\documentclass[twocolumn,floatfix,amsmath,amssymb,10pt]{revtex4}
\usepackage{epsfig}
\usepackage{epstopdf}
\usepackage{epsfig,color}
\usepackage{amsmath}
\usepackage{amsfonts}
\usepackage{amssymb}
\usepackage{float}
\usepackage{latexsym}
\usepackage{dcolumn}
\usepackage{bm}
\usepackage[dvipsnames]{xcolor}
\usepackage{graphicx}

\newcommand{\be}{\begin{equation}}
\newcommand{\ee}{\end{equation}}
\newcommand{\bea}{\begin{eqnarray}}  
\newcommand{\eea}{\end{eqnarray}}

\newcommand{\p}{\partial}
\newcommand{\s}{\sigma}

\newcommand{\la}{\langle}
\newcommand{\ra}{\rangle}
\newcommand{\rd}{\mbox{d}}
\newcommand{\ri}{\mbox{i}}

\newcommand{\eps}{\epsilon}
\newcommand{\nn}{\nonumber}

\newcommand{\vare}{\varepsilon}

\newcommand{\vsigma}{\mbox{\boldmath $\sigma$}}

\newcommand{\vrho}{\mbox{\boldmath $\rho$}}

\newcommand{\hsigma}{\hat{\sigma}}

\newcommand{\bcr}{\begin{array}{clcr}}
\newcommand{\ecr}{\end{array}}

\input{epsf}
\begin{document}

\title{\bf SPECTRUM, LIFSHITZ TRANSITIONS AND
ORBITAL CURRENT IN FRUSTRATED FERMIONIC LADDERS WITH A UNIFORM FLUX}

\author{Bachana Beradze$^{1,2}$}
\author{Alexander Nersesyan$^{1,2,3}$} 
\affiliation{\small{\sl $^1$ The Andronikashvili Institute of Physics, 
0177 Tbilisi, Georgia,}}
\affiliation{\small{\sl $^2$ Ilia State University, 0162 Tbilisi, Georgia}}

\affiliation{\small{\sl$^3$ The Abdus Salam International Centre for Theoretical Physics, 34151, Trieste, Italy}}
\date{\today}

\begin{abstract}

Ultracold Fermi gases  with synthetic gauge field represent an excellent platform to study the 
combined effect of lattice frustration and an  effective magnetic flux  close
to one flux quantum per particle.
The minimal theoretical model to accomplish this task is a system of  spinless noninteracting fermions on a 
triangular two-chain flux ladder. In this paper we consider this model  close to half-filling, 
with interchain hopping  amplitudes alternating along the
zigzag bonds of the ladder and being small.  In such setting  qualitative changes in
the ground state properties of the model, most notably the flux-induced topological Lifshitz transitions 
are shifted towards the values of the flux per a diatomic plaquette ($f$) close to the flux quantum ($f\sim 1/2$).
Geometrical frustration of the lattice breaks
the $k \to \pi - k$ particle-hole spectral symmetry  present in  non-frustrated ladders, and leads to
splitting of the degeneracies at metal-insulator transitions.
A remarkable feature of
a translationally invariant  triangular ladder is the appearance of isolated  Dirac node at the Brillouin zone
boundary, rendering the ground state at $f=1/2$  semi-metallic. 
We calculate the flux dependence of the orbital current and identify
Lifshitz critical points by the singularities in this dependence at a constant chemical potential $\mu$
and constant particle density $\rho$. The absence of the particle-hole symmetry in the triangular ladder leads to the
asymmetry between particle ($\rho>1$) and hole ($\rho<1$) doping and to qualitatively different results for 
the phase diagram,  Lifshitz points and the flux dependence of the orbital current
in the settings  $\rho={\rm const}$ and $\mu={\rm const}$.
\end{abstract}
\maketitle

\section{Introduction}

The effects caused by an external magnetic field that quantizes the spectrum of charged carriers in crystalline lattices
has long been 
one of fundamental problems in condensed matter physics \cite{landau,peierls,onsager, wannier,harper,azbel,brown,hofst}. 
The interplay of the applied magnetic field, dynamical correlations between the particles and superimposed geometrical 
lattice frustration has become the subject of intense research in recent years. Main objects of these studies are Fermi and Bose 
gases on ladders, which are minimal quasi-one-dimensional structures 
that can accommodate  finite magnetic fluxes through elementary plaquettes. 
The kinetic frustration caused by the magnetic field,
as well as the geometrical frustration present in structures like triangular and Creutz ladders,
dramatically change the properties of a system when the flux per particle is close to one (more generally, rational fraction of) flux quantum \cite{hasegawa, barford}. In naturally existing 
quasi-one-dimensional electron systems  so strong fields could hardly be achieved. 
Moreover, in such objects the possibility to focus on purely orbital effects of the magnetic flux
would be obscured by Zeeman splitting of the electron spin states.
Fortunately, ultracold atoms  represent an excellent opportunity to circumvent these limitations.
It has been demonstrated \cite{dalibard} that, in systems of neutral atoms on optical lattices,
Raman-assisted tunneling in two-dimensional optical lattices  generates synthetic gauge fields and thus makes
high effective fluxes feasible.
Moreover, lattice connectivity in optical multi-leg flux ladders can be ensured by
an extra synthetic dimensions which  can be engineered taking advantage of the internal atomic degrees of 
freedom \cite{celi, galitski}. 

The properties of bosonic and fermionic flux ladders have been the subject of many theoretical
and experimental studies \cite{orignac,miyake,atala,livi,NCN-2005,CNN-2006,budich,barbarino2,barbarino1,strinati,caza}.
Ground state phases and their topological properties, chiral boundary currents and
topological Lifshitz transitions taking place in ladders on varying the flux have been investigated.
Lifshitz transitions connect phases with different
numbers of Fermi points in the band spectrum 
and reveal themselves through the singularities
in the flux dependence of the persistent orbital current and its derivative --  the charge stiffness. The existence of cusps in this
dependence for a half-filled standard (square) ladder was indicated in Refs.\cite{NCN-2005,CNN-2006} and more recently
in Refs.\cite{budich,caza}.
Away from half-filling, the behavior of the orbital current as a function of the flux
essentially depends on which quantity, the particle density $\rho =N_f/N$ (canonical ensemble) or the chemical 
potential $\mu$ (grand canonical ensemble),
is chosen to be a fixed parameter \cite{caza}. In standard experiments with ultracold atoms the density is fixed. However, interesting proposals
to control the chemical potential in such systems have recently been reported. These include splitting of two-dimensional ultracold atom systems 
into a physical sample of a homogeneous density and a dilute reservoir
using advances in quantum gas microscopy \cite{mazurenko}, realization of two-terminal configurations based on holographic
techniques for the study of Landauer 
transport \cite{krinner}, and the effective control of chemical potentials  using Rabi coupling between different hyperfine levels 
of the atomic species \cite{lepori}.

In this paper we consider a triangular fermionic ladder which consists of two chains 
coupled by an asymmetric single-particle hopping across zigzag-like interchain links. The  properties of the system are formed due to 
the combined effect of geometrical frustration of the lattice and magnetic flux. 
Here we will resort to a model of noninteracting fermions postponing the study
of correlation effects until a separate forthcoming publication \cite{cooper}.
To extract purely orbital effects of the flux we assume that the particles are  spinless. 
In a standard ladder, 
even in the presence of the flux
the two-band spectrum  maintains a  $k \to \pi - k$ perfect nesting property. 
On the contrary, in a triangular ladder geometrical frustration entails
a dispersive character of interchain hopping, and the latter in turn leads to the breakdown of the 
particle-hole (\emph{ph}) symmetry.
For a weak interchain coupling, the difference between the two ladders
is most pronounced when the flux per particle is close to  flux quantum. The spectrum of the  triangular ladder acquires a new feature: the appearance of two groups of Dirac-like low-energy excitations with different masses.
In particular, in a symmetric (translationally invariant)   triangular ladder   one of the mass gaps vanishes, and
the spectrum exhibits an isolated  Dirac node at the Brillouin zone
boundary.  Close to half-filling this leads to a qualitative difference between the ground states of the standard and symmetric 
triangular ladders: the former is insulating while the latter represents a semi-metal susceptible to quantum dynamical fluctuations.
The absence of \emph{ph}-symmetry in the  triangular ladder to the
asymmetry between particle and hole doping and to qualitatively different results for 
the phase diagram,  Lifshitz points and the flux dependence of the orbital current
in the settings with $\rho={\rm const}$ and $\mu={\rm const}$.

The paper is organized as follows. In Sec.\ref{spec} we introduce the triangular   flux-ladder (TFL) model in the absence of interactions
and discuss its spectral properties assuming
that interchain single-particle hopping is weak. We pay special attention to those features of the model which are
caused by geometrical frustration of the lattice. 
For weakly coupled chains, the representation of the low-energy part of the spectrum in terms of two groups
of massive Dirac fermion proves efficient for the study of topological Lifshitz transitions
for the flux is close to a flux quantum per plaquette.
In Sec.\ref{curr} we provide gauge invariant definitions of the total and relative currents, give general expressions
for their ground-state expectation values  and fully characterize their symmetry properties.
In Sec.\ref{mu-const} we calculate the orbital current at a fixed chemical potential ($\mu=0$).  We show that
the difference between the Dirac masses, caused by the lattice geometrical frustration,
leads to a sequence of two commensurate-incommensurate transitions
showing up as universal square-root singularities in the flux dependence of the current.
In Sec.\ref{rho-const} we describe Lifshitz transitions between various ground-state phases 
at a constant particle density $\rho$ assuming that the value of the flux per plaquette is close to one flux quantum.
We separately consider the cases of hole ($\rho<1$) and particle ($\rho > 1$) doping, identify Lifshitz transitions by the cusps in
the flux dependence of the orbital current 
and emphasize  a qualitative difference between the phase diagrams in the two cases. The summary of the obtained results
including suggestions of their further generalizations is compiled in Sec.\ref{concl}.

\section{Triangular ladder}\label{spec}

\begin{figure}[H]
\centering
\includegraphics[width=3in]{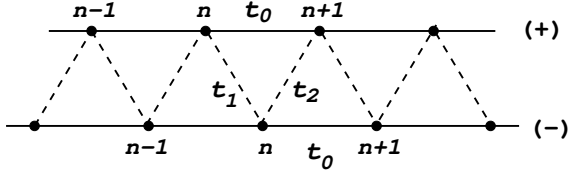}
\caption{\footnotesize Triangular ladder. The integer $n$ labels the diatomic unit cells. $t_0$, $t_1$ and $t_2$ stand for the amplitudes of single-particle nearest-neighbor hopping 
along the chains and between them.} 
\label{zigzag}
\end{figure}

A frustrated ladder with a zigzag geometry of interchain links is shown in Fig.\ref{zigzag}. 
We use the longitudinal gauge in which the flux through the Peierls substitution affects only
hopping along the chains.
The case $t_1 = t_2$ corresponds to a translationally invariant 
ladder.
When one of the two amplitudes,  $t_1$ or $t_2$, vanishes, 
the triangular ladder becomes topologically equivalent to  a standard (rectangular) ladder. 
In the absence of interactions the Hamiltonian of the TFL
reads
\bea
H &=& -~ t_0 \sum_n \sum_{\s=\pm}\left( c^{\dagger}_{n\s} c_{n+1,\s} e^{-i \s\pi  f}+ h.c.  \right)\nn\\
&-& \sum_n \left( t_1 c^{\dagger}_{n+} c_{n-} + t_2 c^{\dagger}_{n+} c_{n-1,-}
+ h.c. \right)
\label{H-f}
\eea
\noindent
Here $c_{n\s}$ and $c^{\dagger}_{n\s}$ are second-quantized operators for a spinless fermion
on the lattice site $(n,\s)$,
the index $\s=\pm$ ~labels the chains, and
$f = \Phi_{\Box}/\phi_0$ is the flux per the diatomic plaquette in units of the flux quantum
$\phi_0 = hc/e$. In momentum representation 
\bea
 H
= \sum_{k} \psi^{\dagger}_k \hat{\cal H}(k) \psi_k, ~~~~
\psi_k = \left(
\begin{array}{clcr}
c_{k+}\\
c_{k-}
\end{array}
\right)
\eea
Here
\bea
\hat{\cal H}(k) = \eps(k) - {\bf h}(k) \cdot \hat{\vsigma} \label{h00-k-matrix}
\eea
is a 2$\times$2 matrix form of the first quantized Hamiltonian whose scalar and vector parts are given by
\bea 
&& \eps(k) = - 2t_0 \cos (\pi f) \cos k, ~~~\nn\\
&& {\bf h} (k) = \{ t_1 + t_2 \cos k, t_2 \sin k, 2t_0 \sin (\pi f) \sin k \}
\label{eps-h}~~~~
\eea
The quasi-momentum $k$, measured in units of inverse lattice spacing $1/a_0$,
varies within the Brillouin zone, $|k|<\pi$, and the Pauli matrices
$\hat{\s}_a ~(a=1,2,3)$  act in the two-dimensional space of the Nambu
spinor $\psi_k$.
The spectrum of $H_0$ consists of two bands labeled by the index  $s=\pm$:
\bea
E_{s} (k) 
= - 2t_0 \cos (\pi f) \cos k + s E_k, ~~
E_k = |{\bf  h}(k)|
\label{spec-E1}~~
\eea
Throughout this paper we assume that the
interchain tunneling amplitudes are positive and small,  $0 < t_{1,2} \ll t_0$.
Moreover, we restrict consideration to the case when the particle density
$
\rho = {N}^{-1} \sum_{n\s} \la c^{\dagger}_{n\s} c_{n\s} \ra
$
is equal or close  to 1. Accordingly the chemical potential is small, 
$|\mu| \ll 2t_0$.  
Without loss of generality, everywhere below we will assume that $t_1, t_2 > 0$.

In the standard ladder  ($t_1 t_2 = 0$), for any $f$ the band spectrum (\ref{spec-E1})  has
the property \cite{NCN-2005,CNN-2006}:
\be
E_{s} (k + \pi) = - E_{-s} (k) \label{ph-band}
\ee
This symmetry is generated by a staggered particle-hole (\emph{ph}) transformation of the fermionic operators,
$c_{k\s} \to c^{\dagger}_{k+\pi, -\s}$.
As a result, if the particle density is fixed at
$\rho = 1$,  Eq.(\ref{ph-band}) necessarily leads to a half-filled band spectrum and vanishing chemical potential, and vice versa. 
The spectral properties of the standard ladder  at $\rho < 1$ and $\rho > 1$ 
(equivalently, for the values of the chemical potential $\pm \mu$)
are equivalent.
In the metallic
phase of a half-filled ladder,  the Fermi momenta at any $f$  form pairs $\pm (k_F, \pi - k_F)$.

As follows from (\ref{spec-E1}),
this is not the case for a TFL
($t_1 t_2 \neq 0$), where geometrical lattice frustration 
makes the band splitting momentum dependent.
In Eq.(\ref{h00-k-matrix}) the "magnetic field" 
$|{\bf h} (k)|$ is not invariant under the transformation $k \to \pi \pm  k$, and 
the perfect nesting property (\ref{ph-band}) is absent.
As a consequence, for a general set of the parameters of the TFL, 
if the particle density is fixed at $\rho =1$, the corresponding value of the chemical
potential $\mu \neq 0$; inversely, if 
$\mu = 0$, then 
$\rho \neq 1$. The properties of the TFL  at $\rho<1$ 
and $\rho>1$ are different.
For  a special value of the flux, $f=1/2$, the Hamiltonian takes the Bloch form
$\hat{\cal H}_k = - {\bf h}(k) \cdot \vsigma$ which implies the $(E,-E)$
symmetry of the band spectrum. Nevertheless the $k \to \pi - k$ 
symmetry is absent at $f=1/2$ as well.

At weak interchain tunneling
the loss of the \emph{ph}-symmetry and
all qualitative changes in structure of the spectrum 
are most
pronounced in the limit $|t_1 - t_2| \ll t_1 + t_2$ and 
for the values of the flux $f$ close to 1/2.
Using the parametrization
\be
f = \frac{1}{2} - \frac{\kappa}{2\pi t_0}, ~~~~|\kappa| \ll 2t_0 \label{f-param}
\ee
and replacing (\ref{spec-E1}) by the approximate expression 
\bea
&& E_{\pm} (k) \simeq - \kappa \cos k\nn\\ 
&&\pm\sqrt{4t^2 _0 \sin^2 k + t^2 _1 + t^2 _2 + 2t_1 t_2 \cos k}
+ O(\kappa^2) \label{zigzag-spec-approx}~~~
\eea
we find that the low-energy states of the spectrum, $|E_{\pm} (k)| \ll 2t_0$,
are located close to the points $k=0$ and $k=\pi$, where excitations 
are described in terms of two groups of "massive Dirac fermions" 
\bea
|k| \ll 1: E_{\pm} (k) \simeq - \kappa \pm \sqrt{k^2 v^2 _F + M^2}\label{zig-0}~~~~\\
|k - \pi| \ll 1: E_{\pm} (k) \simeq \kappa \pm \sqrt{(k - \pi)^2 v^2 _F + m^2}
\label{zig-pi}~~~
\eea
Here $M = t_1 + t_2 $, $m = t_1 - t_2 $ and $v_F = 2t_0  a_0$ is the Fermi velocity (here we drop small corrections to $v_F$ 
of the order $O(t^2_{1,2})$).  
Both masses are  small compared to $2t_0$.

\begin{figure}[H]
\centering
\includegraphics[width=2.3in]{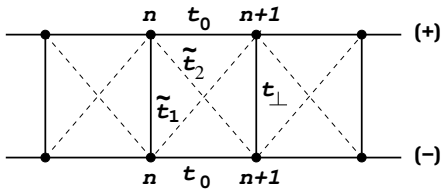}
\caption{\footnotesize Asymmetric Creutz ladder. } 
\label{creutz}
\end{figure}

In fact, the same type of the spectrum characterizes a family of frustrated two-leg ladders encapsulated
in the asymmetric Creutz model, 
shown in Fig.\ref{creutz}. Apart from
the standard on-rung interchain tunneling
($t_{\perp}$), this model includes hopping across the diagonals of elementary plaquettes,
parametrized by two independent amplitudes $\tilde{t}_1$ and $\tilde{t}_2$.  
The case $\tilde{t}_1 = \tilde{t}_2 = 0$, $t_{\perp} \neq 0$ corresponds to a standard (non-frustrated)
ladder.
At $\tilde{t}_1 = \tilde{t}_2 \neq 0$ Fig.\ref{creutz}
represents
a symmetric Creutz ladder \cite{creutz}. At $\tilde{t}_2 = 0$
one of the diagonals of the plaquette is disabled and the lattice becomes topologically
equivalent to a triangular ladder with alternating interchain tunneling amplitudes
$t_{\perp}$ and $\tilde{t}_1$;  in the notations of the model (\ref{zigzag}) the latter coincide with
$t_1$ and $t_2$, respectively.

In the presence of the flux $f$ the Hamiltonian $\hat{\cal H}(k)$ of the asymmetric Creutz ladder
has the form (\ref{h00-k-matrix}) with
\[
{\bf h} (k) = \{t_{\perp} + (\tilde{t}_{1} +\tilde{t}_{2}) \cos k, 
(\tilde{t}_{1} - \tilde{t}_{2}) \sin k,  2t_0 \sin(\pi f) \sin k
\}~~~
\]
The low-energy part of its spectrum is given by formulas (\ref{zig-0}), (\ref{zig-pi}) with the Dirac masses
$M = t_{\perp} + \tilde{t}_1 + \tilde{t}_2$ and $m = t_{\perp} -\tilde{t}_1 - \tilde{t}_2$.
The asymmetry parameter $\tilde{t}_1 - \tilde{t}_2$ is the measure of geometrical frustration of the model of Fig.\ref{creutz}.
The TFL  is the maximally frustrated member of this family. In the rest of this paper we will only deal with
the TFL.

\begin{figure}[H]
\centering
\includegraphics[width=3.3in]{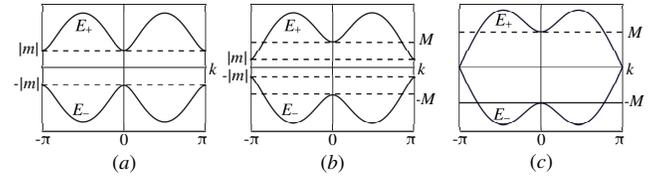}
\caption{\footnotesize Band spectrum at $f=1/2$ a) in the standard ladder, 
b) in the triangular ladder at $m\neq 0$, c) in the triangular ladder at $m=0$.}
\label{spectra-slzl}
\end{figure}

Turning back to the model (\ref{h00-k-matrix})--(\ref{eps-h}), we
choose for certainty $t_1 \geq  t_2 \geq 0$ and introduce a frustration parameter
\be
\delta = \frac{t_2}{t_1} = \frac{M-m}{M+m}, ~~~~~0 \leq \delta \leq 1  \label{frustr-delta}
\ee
In the standard ladder ($\delta = 0$), the $k\to k+\pi $ symmetry keeps the masses equal; Fig.\ref{spectra-slzl}a. 
Geometrical frustration ($\delta \neq 0$) 
makes the Dirac masses different: 
 see  Fig.\ref{spectra-slzl}b. 
The difference is most striking in the
translationally invariant case ($m\to 0$), 
when  the  degree of frustration reaches its maximum, $\delta \to 1$;
Fig.\ref{spectra-slzl}c.
While at $f=1/2$ ($\kappa = 0$), the ground state of a half-filled
standard ladder is insulating, under the same conditions the presence of 
a Dirac node at $k=\pi$ 
renders the spectrum of the symmetric TFL gapless \footnote{In fact, the location of the lowest-energy states close to the zone boundary
$k=\pi$ is a robust property of the model valid as long as $|t_1 - t_2| \ll |t_1 + t_2|$  irrespective of the magnitude of the ratios $t_{1,2}/t_0$. On the other hand, a local minimum of $E_k$ at $k=0$ only exists under the condition $t_1 t_2 < 4t^2 _0$; otherwise $E_k$ has a local
maximum at $k=0$.}
(Fig.\ref{spectra-slzl}c).

The appearance of the parameter $\kappa$ in the expressions  (\ref{zig-0}), (\ref{zig-pi}) makes
the spectrum deformable: when $\kappa$ is varied,
the parts of  the dispersion curves at  $k\sim 0$ and $k \sim \pi$ 
"move" along the energy axis in
opposite directions.
This fact explains the existence of flux induced topological Lifshitz transitions \cite{volovik} between ground state phases with
different numbers of Fermi points. Moreover, it also indicates that, as long as $|\mu|, M, |m| \ll 2t_0$, all these transitions 
take place in the vicinity of the point $f=1/2$, i.e. at small $\kappa$.
The  difference between the mass gaps  $M$ and $|m|$
affects the ground state phase diagram  and the sequence
of Lifshitz transitions in a TFL  even at $\mu = 0$. The transitions show up
in the experimentally observable singularities in the flux dependence of the orbital current and the anomalies
of the charge stiffness of the system. 
A discussion on this issue in the context of square flux ladders, together with relevant references to 
ongoing experimental work, can be found in the recent publication \cite{caza}.

To complete this section, let us
 briefly comment on topological properties of the insulating phases of the noninteracting TFL. 
For instance, at $f=1/2$ such phases are realized at $m \neq 0$ under the condition that $-|m| < \mu < |m|$. With $M>0$
there are two insulating phases corresponding to opposite signs of the mass $m$ (of course, the thermodynamical properties depend only
on $|m|$). The topological properties of a gapped ground state 
can be  inferred
from the 2$\times$2 matrix structure of the Bloch Hamiltonian $\hat{\cal H}(k) = - {\bf h}(k) \cdot \hat{\vsigma}$ in Eq.(\ref{h00-k-matrix}). 
In spite of the different structure of the Nambu spinors,
the first quantized Hamiltonian $\hat{\cal H}(k)$ of the TFL  and that of the 1D spinless
p-wave superconductor (Kitaev's model) \cite{kitaev-toy,alicea}
 have the same matrix form
and same symmetry properties of the "magnetic" field ${\bf h}(k)$
under transformation $k \to - k$. According to conventional definitions (see e.g. Ref. \onlinecite{ludwig}), a matrix
Hamiltonian ${\cal H}(k)$
possesses time reversal (${\cal T}$), charge-conjugation (${\cal C}$) and chiral (${\cal S}$) symmetry if the following conditions 
\bea
U_{\cal T} {\cal H}^* (k) U^{-1}_{\cal T}  &=&  {\cal H}(-k) \nn\\
U_{\cal C} {\cal H}^* (k) U^{-1}_{\cal C}  &=&  - {\cal H}(-k) \nn\\
U_{\cal S} {\cal H}(k) U^{-1}_{\cal S}  &=&  - {\cal H}(k) \nn
\eea
are satisfied (here the $U$s are unitary matrices). Apparently, charge conjugation symmetry is present
(with $U_{\cal C} = \s_3$), whereas time reversal and chiral symmetries are
explicitly broken. 
These facts place the TFL  into the symmetry class D \cite{ludwig}.
In this class,  maps of the Brillouin zone to the sphere traversed by  a unit vector ${\bf h}(k)/|{\bf h}(k)|$ form a
$\mathbb{Z}_2$ group. Following the same arguments as those in \cite{alicea}, one finds out that homotopically nontrivial maps take place at
$m<0$ (more generally at $Mm < 0$). Thus the insulating ground state of the TFL  at $m<0$ is topologically nontrivial, characterized
by a nonzero Zak phase $\phi_{\rm Zak} = \pi$, whereas the phase at $m>0$ is nontopological ($\phi_{\rm Zak} =0$).
It is clear that the insulating phase of a standard ladder with $M=m$ is topologically trivial.

\section{Orbital current}\label{curr}

The gauge invariant  current  through a directed link $\la n,n+1 \ra$ of the chain   $\s$
is defined as
\bea
{J}^{\s}_{n,n+1} = - \ri t_0  c^{\dagger}_{n\s} c_{n+1,\s} e^{- i \pi  \s f}  +   h.c.,
~~~~~(\s=\pm) \nn
\eea
The second quantized total and relative current operators are given by 
\bea
{J}_{\rm tot} &=& N^{-1}\sum_{n\s}  {J}^{\s}_{n,n+1}
= N^{-1}\sum_k \psi^{\dagger}_k \hat{J}_{\rm tot} (k) \psi_k~~
\label{total}\\
{J}_{\rm rel} &=& N^{-1} \sum_{n\s}  \s {J}^{\s}_{n,n+1}
= N^{-1}\sum_k \psi^{\dagger}_k \hat{J}_{\rm rel} (k) \psi_k~~
\label{relat}
\eea
where 
\bea
 \hat{J}_{\rm tot} (k) &=& 2t_0 
\left[\cos (\pi f) \sin k - \hat{\s}_3 \sin (\pi f) \cos k  \right]~~~
\label{tot-k}\\
\hat{J}_{\rm rel} (k) &=& 2t_0 
\left[\hat{\s}_3 \cos (\pi f) \sin k -  \sin (\pi f) \cos k  \right]~~~
\label{rel-k}
\eea
are first quantized 2$\times$2 matrix versions of the currents in the chain basis.
\medskip

\begin{figure}[H]
\centering
\includegraphics[width=2.2in]{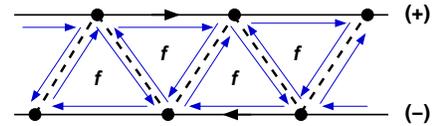}
\caption{\footnotesize Right and left fermions propagate along the upper $(\s=+)$
and lower $(\s=-)$ chains, respectively. Net local currents along the zigzag links
vanish.
} 
\label{spat-separA}
\end{figure}

The Hamiltonian (\ref{h00-k-matrix})
is invariant under parity transformation $P$ ($k\to -k$)
combined with the interchange of the two chains of the ladder
$P_{12}$ ($\s \to -\s$):
\be
{\cal P} = PP_{12},~~~{\cal P} \hat{\cal H}(k)  {\cal P}^{-1} =
\hat{\s}_1 \hat{\cal H}(-k) \hsigma_1 = \hat{\cal H}(k)
\label{pp12}
\ee
Under ${\cal P}$  the total current $ \hat{J}_{\rm tot} (k)$ changes its sign.
Therefore, as long as this symmetry remain unbroken, which is an obvious case for the
noninteracting model, the  expectation value  of the total current vanishes.
This conclusioin can be changed in strongly correlated phases of interacting fermions in
which  ${\cal P}$-symmetry may be spontaneously broken \cite{cooper}.
On the other hand, the operator 
$ \hat{J}_{\rm rel} (k)$ 
remains invariant under ${\cal P}$ and, hence, except for special values of $f$,  
its expectation value is  nonzero. Thus, at $f\neq 0$ the ground state of the TFL is characterized
by a nonzero orbital 
current $\la \hat{J}_{\rm rel} \ra$ flowing in opposite directions on the upper
and lower chains (Fig.\ref{spat-separA}).
\medskip

The ground state expectation value of the orbital current 
is given by
\bea
&&{\cal J}(f) \equiv \la J_{\rm rel} \ra~~~\nn\\
&&= \frac{2t_0}{N} \sum_{k\s}
\la c^{\dagger}_{k\s} c_{k\s}  \ra
\left[
\s \sin k \cos (\pi f) -  \cos k \sin (\pi f) 
\right]
\label{rel-j-kA}~~~~~~
\eea
The occupation numbers $\la c^{\dagger}_{k\s} c_{k\s}  \ra$
can be easily computed using
the Matsubara Green's function
in the 2$\times$2 matrix representation, $\hat{G}(k,\vare) = \left( \ri \vare + \mu -  \hat{\cal H}_k  \right)^{-1}$,
with $ \hat{\cal H}_k  $ given by (\ref{h00-k-matrix}).  The result is
\bea
{\cal J}(f) 
=  -  \frac{2t_0}{\pi}\int_{0}^{\pi} \rd k ~{\cal F}(k;f),\nn\\
 {\cal F}(k;f) =\Big\{
\frac{t_0 \sin(2\pi f)\sin^2 k }{E_k} 
\times [n_+ (k) - n_- (k)]\nn\\
+ \cos k \sin (\pi f)  \left[ n_+ (k) + n_- (k) \right]
\Big\}
\label{Jrel-int}~~~~~~~
\eea
where $E_k$ is given by (\ref{spec-E1})
and $n_{\pm} (k) =  \theta \left[ \mu - E_{\pm} (k)  \right]$
are the ground state Fermi distribution functions for the upper and lower bands, $\theta (x)$
being the Heaviside step function. As follows from (\ref{Jrel-int}), the orbital current  is contributed by
\emph{all} occupied state of the two-band spectrum and, as has been mentioned earlier \cite{NCN-2005, CNN-2006},
does not represent an infrared phenomenon.

For arbitrary values of the chemical potential $\mu$ the current vanishes
at integer values of the flux, $f = 0,\pm 1, \pm 2, \ldots$, and has the periodicity
property\\ ${\cal J} (f+2) = {\cal J}(f)$. At the special value 
$f=1/2$
\bea
{\cal J} (1/2)=  - \frac{2t_0}{\pi} \int_{0}^{\pi} \rd k ~
\cos k  ~ \left[ n_+ (k) + n_- (k) \right]
\label{J-1/2A}
\eea
When $m \neq 0$ and $\mu$
lies within the spectral gap (see Fig.\ref{spectra-slzl}b),
$
-|m| \leq \mu \leq |m|,
$
the ground state is insulating with an empty upper band $E^+ (k)$ and fully filled lower band
$E^- (k)$. In this case for all $k$
$n_+ (k) = 0$, $n_- (k) = 1$, and the integral in (\ref{J-1/2A}) vanishes: 
$
{\cal J}(1/2)|_{\rm insul} = 0.
$
This result 
is valid for all $f = l + 1/2$ as long as  the ground state is insulating.
If $\mu > |m|$ or $\mu < - |m|$, a "Fermi surface" appears and the ground state becomes metallic
with a finite $k_F$, in which case ${\cal J}(1/2) \sim \sin k_F \neq 0$. 

Shifting the flux by one flux quantum, $f \to f + 1$, or inverting it with respect to the point 1/2,
$f \to 1 - f$, 
generates a non-staggered \emph{ph}-transformation of the band spectrum,
$
E_s (k, 1\pm f) = - E_{-s} (k,f). 
$
Accordingly 
$
n_{\pm}(k; 1-f, \mu) = 1 - n_{\mp} (k; f, - \mu). 
$
Then from the definition (\ref{Jrel-int}) it follows that
\bea
{\cal J} (f+1; \mu) &=&  {\cal J}(f; - \mu) \label{+sym}\\
{\cal J} (1-f; \mu) &=& - {\cal J}(f; - \mu) \label{-sym}
\eea
When the current is defined at a fixed density $\rho$,
the above transformation 
should be replaced by 
\bea
{\cal J} (f+1; \rho) &=&  {\cal J}(f; 2-\rho) \label{+symA}\\
{\cal J} (1-f; \rho) &=& - {\cal J}(f; 2-\rho) \label{-symA}
\eea
Thus, in the TFL, the orbital current is symmetric (antisymmetric) under 
transformations $f \to f+1$
($f \to 1 - f$)  
combined with  particle-hole transformations $\mu \to - \mu$ or $\rho \to 2 - \rho$.  For comparison, for a 
standard ladder, the spectral
property (\ref{ph-band}) 
makes the current symmetric under the change $\mu\to - \mu$, 
and ${\cal J} (f) $  is a periodic function of the flux with the period $\Delta f = 1$.

\section{Commensurate-incommensurate transitions at constant chemical potential}\label{mu-const}

At $f \ll 1$ the current is proportional to the flux and the difference between the standard and
triangular ladders is not significant. 
Below we  concentrate on the region $|f-1/2| \ll 1$. 
Here we will assume that the TFL is in contact with a reservoir which supports a fixed value of the chemical potential
$\mu=0$.  This value of $\mu$ is chosen only for the sake of simplicity of the discussion.
We will show that geometrical  frustration of the TFL 
splits the Lifshitz (metal-insulator) transition of the square ladder into two consecutive 
transitions taking place at $\kappa = |m|$ and $\kappa = M$. In the vicinity of these transitions  the current 
is a nonanalytic function of the flux.
 \medskip
 
 \begin{figure} [H]
\centering
\includegraphics[width=3.35in]{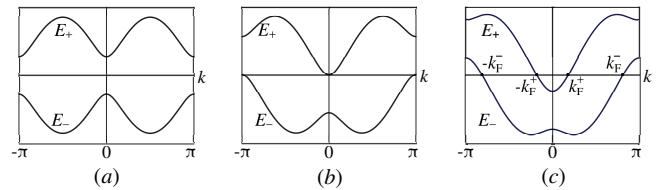}
\caption{\footnotesize Commensurate-incommensurate transition in a standard ladder.
a) Insulating phase at $\kappa < |m|$; b) critical point $\kappa = |m|$;
c) metallic phase at $\kappa > |m|$.
} 
\label{stand-CIC}
\end{figure}
 
Using (\ref{f-param}) 
and the expression (\ref{zigzag-spec-approx})  valid at small $\kappa$,
we represent the current as
\bea
&& {\cal J}(\kappa) = - \frac{1}{\pi} \sum_{s=\pm}
\int_0 ^{\pi} \rd k~\Phi_{s} (k) n_s (k) ,  \label{J-small-kappa} \\ 
&&~~~~\Phi_{s} (k) = W \cos k 
+ s
 \frac{\kappa \sin^2 k }{{\cal E}_k  }, \nn\\
 && {\cal E} (k) =\sqrt{\sin^2 k +  \tau^2 _1 + \tau^2 _2 + 2 \tau_1 \tau_2 \cos k} 
 \nn
 \eea
where $W = 2t_0$ and  $\tau_{1,2} = t_{1,2}/W\ll 1$.

\noindent
We start with the insulating phase 
at $m \neq 0$, Fig.\ref{spectra-slzl}b, 
realized at $\kappa < |m|$.
For this phase 
$n_+ (k) = 0$, 
$n_- (k) = 1$ for all $k$, 
and the orbital current is a linear function of $\kappa$. With the logarithmic accuracy
\bea
{\cal J}_0 (\kappa)  &=& \frac{C_0}{\pi} \kappa,\label{J-insul-phase}\\
C_0 &=& \int_0 ^{\pi} \rd k~ \frac{\sin^2 k}{{\cal E}_k}\nn\\
&=& 2 - \frac{1}{2W^2 } \left( M^2 \ln \frac{W}{M} + m^2   \ln \frac{W}{|m|} \right)\nn
\eea

For the insulating phase  of the  standard ladder the result is not much different: one only has to set in (\ref{J-insul-phase})
$M = m$. The difference emerges when the transitions to metallic phases are compared.  For a half-filled standard ladder, the
\emph{ph}-symmetry (\ref{ph-band})
 implies that  if for some $\kappa$ $E_+ (0) = 0$, then necessarily $E_- (\pi) = 0$. 
Fig.\ref{stand-CIC}b shows the spectrum  at the doubly degenerate critical point
$\kappa = m =t_1$. As follows from  Fig.\ref{stand-CIC}c, at $\kappa > m$ 
the ground state becomes metallic with
\emph{four} Fermi points, $ \pm k_0$, and $ \pm (\pi - k_0)$.  
In this phase, close to the critical point
$
k_0 = \sqrt{\kappa^2 - m^2} / v_F $. 
One then finds that
\bea
{\cal J} (\kappa) = {\cal J}_0 (\kappa)  + {\cal J} _{\rm sing}(\kappa) 
\label{curr-sum}
\eea
where the singular part of the current 
displays a  square-root dependence on $\kappa$, typical for a quantum commensurate-incommensurate (C-IC)
transition \cite{JN,PT}:
\bea
 {\cal J} _{\rm sing}(\kappa)  = - \frac{W}{\pi} 
 \left( \int_0 ^{k_0} + \int_{0}^{\pi - k_0}  \right) \rd k~\cos k \nn\\
 = - \frac{2}{\pi} \sqrt{\kappa^2 - m^2}  + O [(\kappa^2 - m^2 )^{3/2}]\label{SL-sing-J}
\eea

\begin{figure}[H]
\centering
\includegraphics[width=3in]{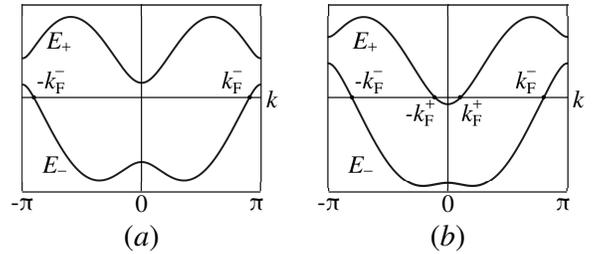}
\caption{\footnotesize 
Commensurate-incommensurate transition in a TFL.
a) 2FP metallic phase at $|m| < \kappa < M$;
b) 4FP metallic phase at $\kappa > M$. 
} 
\label{ZL-CIC}
\end{figure}

As already mentioned, in the TFL, an arbitrarily weak frustration ($\delta \ll 1$)
removes the degeneracy of the two mass gaps, which leads to
two Lifshitz transitions.
Increasing $\kappa$ in the insulating phase, Fig.\ref{spectra-slzl}b, one first observes a transition at $\kappa = |m|$, 
where $E_- (\pi; |m|) = 0$ with  $E_+ (0; |m|) > 0$.
This is a transition 
to a metallic phase with \emph{two} Fermi points, shown in Fig.\ref {ZL-CIC}a.
This phase is realized in the region $|m| < \kappa < M$. 
Close to the critical point $0 < \kappa - |m|\ll |m|$ one obtains
${\cal J}_{\rm sing} (\kappa) \simeq \pi^{-1} \sqrt{\kappa^2 - m^2}$.
Further increasing $\kappa$ leads to the second Lifshitz transition 
to a metallic phase with \emph{four} Fermi points, Fig. \ref {ZL-CIC}b.
This transition occurs at $\kappa = M$ where $E_+(0; M) = 0$.
One then finds that ${\cal J}_{\rm sing} (\kappa) \simeq  - {\pi}^{-1} \sqrt{\kappa^2 - M^2 }$ at
$0 < \kappa - M \ll M$.
At larger values of $\kappa$, namely at $M \ll \kappa \ll W \ln^{-1/2} (W/M)$, the current 
is easily shown to be inversely proportional to $\kappa$:
$
{\cal J} (\kappa) \simeq (M^2 + m^2)/2\pi \kappa.
$

\begin{figure}[H]
\centering
\includegraphics[width=2.9in]{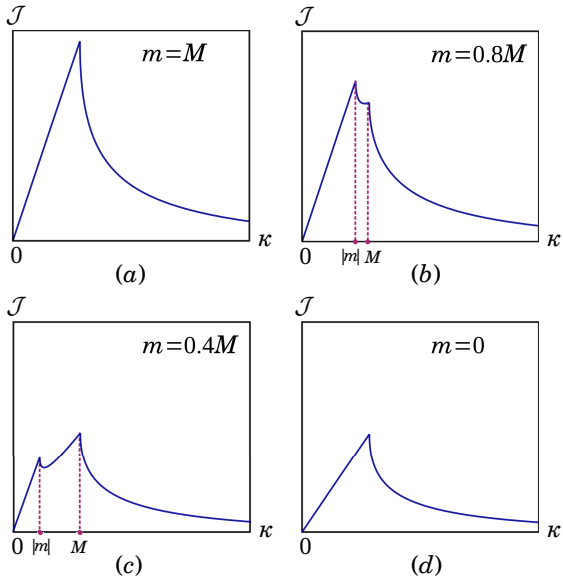}
\caption{\footnotesize Orbital current at a fixed chemical potential $\mu=0$ and small
$\kappa$ for different values of $|m|$, with $M = 0.4 t_0$. (a) A single Lifshitz transition in
a standard ladder, $|m| = M$; (b) and (c) -- a sequence of two transitions  at $|m| < M$; 
(d) a single transition at $\kappa = M$ when $m \to 0$.}
\label{CIC-trans-mu}
\end{figure}

At $m \to 0$ the insulating phase of the TFL disappears, implying that
the transition  at $\kappa = |m|$  disappears as well. The phase at $0<\kappa < M$
is metallic with \emph{two} Fermi points. The current in this region is a linear function of $\kappa$ but
with a different slope as compared to Eq.(\ref{J-insul-phase}):
\be
{\cal J} (\kappa) = \frac{\kappa}{\pi} \left( 1 - \frac{M^2}{2W^2 }  \ln \frac{W}{M}  \right)
\label{m=0-J}
\ee
At  $\kappa = M$ the ladder crosses over to the metallic state with four Fermi points.
The flux dependence of the orbital current at $m \neq 0$ and small $\kappa$ is shown in Figs.\ref{CIC-trans-mu}.
The quantity diverging at the C-IC  transitions 
is the relative charge stiffness
defined as a response function \cite{SS,GS}
\be
D = \frac{1}{2 \phi_0} \frac{\p {\cal J} }{\p f}
= - \frac{\pi t_0}{\phi_0} \frac{\p {\cal J}} {\p \kappa} \label{D-def}
\ee 
In the context of usual C-IC transitions taking place in the  charge or spin excitations
of  various one-dimensional Fermi systems \cite{GNT}, the quantity $D$ is similar to the compressibility or spin susceptibility.
We see that, when the frustration ratio $\delta$,  Eq.(\ref{frustr-delta}), is nonzero,
the single peak of $D$ at the metal-insulator transition of the standard ladder splits into two
peaks that move apart as $\delta$ is increased (Figs.\ref{CIC-trans-mu}).
In the vicinity of these transitions
$D(\kappa)$ displays universal square-root
singularities \cite{JN,PT}:
\be
D (\kappa) \simeq \frac{t_0}{\phi_0} \frac{\kappa_{c}}{\sqrt{\kappa^2 - \kappa^2 _{c}}}
\theta(\kappa^2 - \kappa^2 _c)
\label{D}
\ee
with $\kappa_c = |m|$ or $M$.

Ground state phases and the transitions between them at a
small nonzero $\mu$ can be categorized along the same lines.
At a constant $\mu \neq 0$ 
all Lifshitz transitions accessed by varying the flux
belong to the C-IC  universality class, with a typical behavior of the charge stiffness 
given by (\ref{D}). The reason for that is simple. A generic C-IC transition is a transition from an insulator with a parabolic
two-band spectrum to a metal, taking place when 
varying the chemical potential leads
to the appearance of extra particles(holes) in the conduction (valence) band. 
In the TFL  with a constant $\mu$
the deformation of the
two-band spectrum caused by a small $\kappa$ can be recast as
the appearance of variable effective chemical potentials in the Dirac-like parts of the spectrum
in the regions $k\sim 0$ and $\pi$. Repopulation of these low-energy states in each of these region occurs independently and
at different values of $\kappa$. Therefore, each of these transitions is of the C-IC type.

\section{Phase diagram 
and orbital current at constant density}\label{rho-const}
In this section we  discuss  the effect of the flux on the spectrum, orbital current and phase diagram
of the noninteracting TFL
at a constant density,  fixed at a value close to 1:
$
\rho = 1 + n, ~|n| \ll 1 
$.
Such setup is typical  for ultracold atoms 
on optical lattices.
As before, we assume that  $|f-1/2|\ll 1$.
Except for the special case $\rho=1$, Lifshitz transitions at $\rho=$const are different from those
that occur at  $\mu=$const. 
Moreover, as we show below, contrary to the standard ladder,
in the TFL 
the absence  of \emph{ph}-symmetry 
makes the pictures at 
hole doping ($n < 0$) and particle doping (${n>0}$) 
qualitatively different.

At $\kappa > (M+|m|)/2$ the two bands of the spectrum overlap. If the chemical potential is located
between the energies $E_+ (0)$ and $E_- (\pi)$, i.e.
$M-\kappa < \mu < \kappa - |m|$, the ground state is metallic with four  Fermi points $\pm k_F^+, ~\pm k_F^-$
in the spectrum. We will call this state the 4FP-phase. Imposing an additional restriction upon the density, 
\be
\pi |n| < \frac{\sqrt{M^2 - m^2}}{W} \label{n<}
\ee
we disregard the situation when 
$\mu < E_- (0)$ and all four Fermi points belong to the $E_- (k)$ band.
Starting from the 4FP phase we will
then trace its evolution  as $\kappa$ is decreased.
At a constant density 
$\rho \neq 1$ the Fermi momenta  $k^{\pm}_F$ are subject to the constraint
$
k^+ _F + k^- _F = \pi (1+n).  
$
Since $|\kappa| \ll W$ and $|n| \ll 1$,
$k_F^+$ and $k_F^-$ are close to the points $k=0$ and $\pi$,  respectively, where
the spectrum has the Dirac form  (\ref{zig-0}) and (\ref{zig-pi}). 
Therefore we can set $k^+ _F = p_+, ~k^- _F = \pi - p_- $, where the momenta $p_{\pm}$ are small,
positive and satisfy 
\be
p_+ - p_- = \pi n \label{normal2}
\ee
At given values of the Dirac masses the
dependence of $p_{\pm}$ and $\mu$ on the flux is described by the equations
\be
\sqrt{p^2 _+ v^2 _F + M^2} = \kappa + \mu, ~~~~\sqrt{p^2 _- v^2 _F + m^2} = \kappa - \mu
\label{eqs-basic}
\ee
valid in the 4FP phase for any sign of $n$.

\subsection{$\vrho{\bf <1 ~(n<0)}$}
 \label{r-less-1}

\begin{figure}[H]
\centering
\includegraphics[width=3.4in]{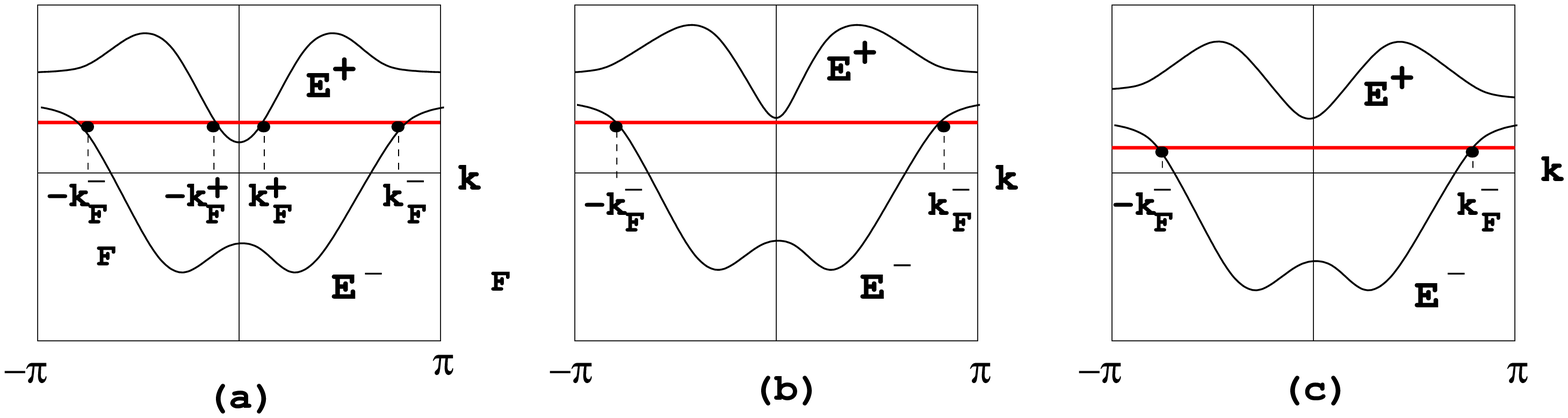}
\caption{\footnotesize Case $\rho<1$. (a) 4FP metallic phase at $\kappa > \kappa_c$; (b) Lifshitz critical point at
$\kappa = \kappa_c$; (c) 2FP metallic phase at $\kappa < \kappa_c .$} 
\label{ro-more-4A}
\end{figure}

According to (\ref{normal2}) at $n<0$ one has $p_+ < p_-$. The momentum $p_+$ decreases with $\kappa$ and vanishes
at some critical value $\kappa = \kappa_c$ where the band $E_+(k)$ gets empty. This is a Lifshitz transition at which  the number of Fermi points changes 
from four to two (Fig.\ref{ro-more-4A}b); we label
the new state as the 2FP phase.  The critical values
of $\kappa$ and $\mu$ are given by 
$
\kappa_{c} =(M + Q_m )/{2}$,~
$\mu_{c}  = (M - Q_m )/{2} $, 
where $Q_m = \sqrt{(\pi n)^2  v^2 _F + m^2}$. 
 As shown below,  the dependence $p_+ = 
p_+ (\kappa)$ at small $\kappa - \kappa_c$ determines the singularities in the flux dependence of the  orbital current
near the transition.

From (\ref{normal2} and (\ref{eqs-basic}) one derives the quadratic equation for $p_+$ valid in the 4FP-phase
\be
Ap^2 _+ + B p_+ - C = 0, ~~~\kappa > \kappa_c\label{quadr-equa}
\ee
where 
\bea
&&A= 1 - \left(\frac{\pi |n|}{2\kappa} \right)^2, ~~
B= \pi |n| \left(1 + \frac{\kappa_c \mu_c}{\kappa^2} \right), \nn\\
&&~~~C = \left(1 - \frac{\mu^2 _c}{\kappa^2}  \right) (\kappa^2 - \kappa^2 _c) 
\label{ABC}
\eea
Obviously $B>0$. Since  $\kappa > \kappa_c > \mu_c$, the coefficient $C$ is positive as well. One also finds that
\[
\frac{\pi |n|}{2\kappa} \leq \frac{\pi |n|}{M + \sqrt{(\pi n)^2 + m^2}} < 1
\]
implying that $A>0$. Focusing on the vicinity of the Lifshitz point we introduce the small parameter \\
$
\zeta = (\kappa - \kappa_c)/\kappa_c$, $0 < \zeta \ll 1.
$
The solution  of Eq.(\ref{quadr-equa}) 
satisfying the boundary condition $p_+ (0) = 0$ reads
\bea
&& p _+ (\zeta) =\frac{M\pi |n|} {A(M+Q_m)}
\left[ \sqrt{1 + (\zeta/\zeta_0)} - 1 \right]  \label{P-int}\\
&&~~~~
\zeta_0 = \frac{1}{2A} \frac{M(\pi n W)^2}{Q_m (M+Q_m)^2}
\nn
\eea
Eq.(\ref{P-int}) is applicable at any $|n| \ll 1$. 
Assuming that $A \lesssim 1$ and $M\gg |m|$ one easily checks that for any $|n|\ll 1$
the parameter $\zeta_0 \ll 1$. Therefore there are two regions of small $\zeta$.
In the immediate vicinity of the transition point, $0 < \zeta \ll \zeta_0$, the dependence $p_+ (\zeta)$
is linear
\be
p _+ =  \frac{1}{v_F}\frac{Q_m (M+Q_m)}{W^2 (\pi |n|)} \zeta
\label{linear-p}
\ee
At larger 
deviations from the criticality, $\zeta_0 \ll \zeta \ll 1$, this dependence crosses over to
the square-root asymptotics
\bea
p _+ = \frac{1}{W} \sqrt{\frac{2 MQ_m \zeta}{A}}
\label{sqrt-zeta}
\eea
typical for a C-IC transition.
In the limit $n \to 0$ ($\rho \to 1^-$)  $\zeta_0 \to 0$ and the square-root dependence becomes exact:
$p_+ = \sqrt{2M|m| \zeta}.$

At $\kappa < \kappa_{c}$ the spectrum describes 
the metallic 2FP phase,
Fig.\ref{ro-more-4A}c, in which $k^-_ F$ 
stays fixed at the value $\pi (1 - |n|)$ and does not depend on $\kappa$. 
As long as  $\rho < 1$,  
the lower band $E_- (k)$
remains partially filled, so  the 2FP phase is stable down to the point
$\kappa = 0~(f=1/2)$. 
Thus, at hole doping 
the ground state of the TFL with $m \neq 0$ always remains metallic.
On decreasing $\kappa$ one finds a single Lifshitz transition separating the 4FP and 2FP phases.

\begin{figure}[H]
\centering
\includegraphics[width=3.3in]{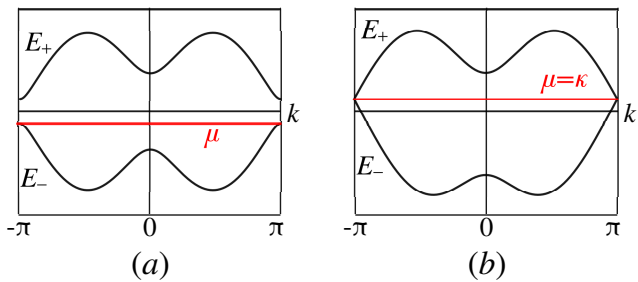}
\caption{\footnotesize $\rho = 1^-$. (a) $m\neq 0$; (b) $m=0$.}
\label{1+A}
\end{figure}

In the 2FP phase the chemical potential is given by
$
\mu (\kappa) = \kappa - \sqrt{(\pi n)^2 W^2 + m^2}.  
$
When $\rho \to 1^-$ ($n \to 0^-$),
$\mu$ reaches the top of the  band $E_- (k)$ (see Fig.\ref{1+A}a)
\be
\lim_{n \to 0} \mu(\kappa) = \kappa - |m|\label{mu-n-0}
\ee
and the latter becomes fully filled. At $m \neq 0$ the upper and lower bands are separated
by a finite gap $2|m|$. 
However at $m = 0$
a Dirac node appears in the spectrum of the $\pi$-fermions at the energy
$E_- (\pi) = \kappa$ and the chemical potential is located just at this energy: $\mu = \kappa$.
This is a semimetallic phase with a single Dirac point and a fully filled "Dirac sea";
see Fig.\ref{1+A}b.

Let us now estimate the orbital current 
at $\rho < 1$.
According to the definition (\ref{J-small-kappa}), in the 4FP phase
the current is a sum of the contributions
of the partially filled upper and lower bands:
\bea
{\cal J}_{\rm 4FP} (\kappa) = {\cal J}_+ (\kappa) + {\cal J}_- (\kappa) 
= - \frac{1}{\pi} \sum_{s=\pm}\int_0 ^{k^s _F} \rd k~\Phi_s (k)~~
\label{4fp-curr}
\eea
where $k^- _F = \pi -p_+$ and $k^+ _F = p_+$
 is given by (\ref{P-int}).
Since $k^+ _F \ll 1$, in the leading order 
\bea
 {\cal J}_+  
  = - \frac{W}{\pi} p_+  + O(p _+ ^3)
\label{J+}
 \eea
 For $ {\cal J}_- $ we have
 \bea
 {\cal J}_- &=&  - \frac{1}{\pi} \left(\int_0 ^{\pi} - \int_{\pi - p_-}^{\pi}  \right)\rd k~
\left(2t_0 \cos k - \frac{\kappa \sin^2 k}{{\cal E}_k}   \right)\nn\\
&\equiv& {\cal J}_0  + {\cal J}_1 
 \eea
Here ${\cal J}_0  = C_0 \kappa /\pi$ given by (\ref{J-insul-phase})
is the current in the band insulator phase, contributed
by the fully filled band $E_- (k)$.
  For the integral ${\cal J}_1 $  we obtain
\bea
{\cal J}_1 
&=& - \frac{W}{\pi} \sin\tilde{p}_+-\frac{C_1 \kappa }{\pi}  \label{J1}\\
C_1 &=&
\frac{{m}^2}{4W^2} \left(\sinh 2\alpha_0 - 2\alpha_0 \right) \nn
\eea
where  $\tilde{p}_+ = p_+ + \pi |n|$, $
\sinh \alpha_0 =  \tilde{p}_+  / |m|$.
So, in the lowest order in the small parameters $k^+_F$ and $|n|$
the current in the 4FP phase can be written as
\bea
{\cal J}_{\rm 4FP} (\kappa) = - W|n| - \frac{2W}{\pi} k^+ _F  + \frac{C_0 - C_1 }{\pi} \kappa, ~\kappa > \kappa_c~~
\label{J-4FP-fin}
\eea
In the 2FP phase 
the upper band does not contribute to the current, 
so the latter 
is obtained
from (\ref{J-4FP-fin}) by setting $p_+  =0$.
At $\kappa < \kappa_c$ the parameter $C_1$ depends only on the density $|n|$
and for any ratio $\pi |n| W / |m| $  is small compared to $C_0$.
Neglecting 
small corrections of the order of $M^2 \ln M$, we can write
\be
{\cal J}_{\rm 2FP} (\kappa) = \frac{2}{\pi} (\kappa - \kappa_0), ~~~\kappa_0 = t_0 \pi |n| ~~~~(\kappa < \kappa_c) \label{J-2FP-fin}
\ee

We  see that, in the 2FP phase, the vacuum current is a linear function of $\kappa$. 
The point where it changes the sign is determined by the hole density $|n|$. The fact that ${\cal J}_{\rm 2FP} (0) \neq 0$ is not unexpected. It reflects the lack of the \emph{ph}-symmetry of the model
and the absence of the $\rho \to 2-\rho$ invariance of the thermodynamic quantities
(see the discussion in Sec.\ref{curr}).

Comparing (\ref{J-4FP-fin}) and (\ref{J-2FP-fin}) and taking into account (\ref{linear-p}) we observe that 
the current is continuous across the 2FP-to-4FP Lifshitz transition but exhibits a cusp at $\kappa = \kappa_c$.
As a result, at the critical point the relative charge stiffness $D(\kappa)$,
Eq.(\ref{D-def}), which is  proportional to the derivative $\p {\cal J}/\p \kappa$, undergoes a jump: 
\bea
&&\frac{\p {\cal J}_{\rm 2FP}}{\p \kappa}\Big|_{\kappa_c}\equiv \lim_{\kappa \to \kappa_c -0}\frac{\p {\cal J}(\kappa)}{\p \kappa} =
 \frac{2}{\pi} \label{deriv-2fp}\\
&&\frac{\p {\cal J}_{\rm 4FP}}{\p \kappa}\Big|_{\kappa_c}  \equiv \lim_{\kappa \to \kappa_c +0}\frac{\p {\cal J}(\kappa)}{\p \kappa} \nn\\
&&\qquad =
\frac{2}{\pi} \left[1 - \sqrt{1 + \left(\frac{m}{\pi n W}\right)^2} \right], ~n\neq 0
\label{deriv-4fp}
\eea
(In deriving formula (\ref{deriv-4fp}) we neglected a small relative correction proportional to
$\p C_1 / \p p_+$.)
Thus, at a constant density but away from half-filling the
square-root dependence of the current on the flux at the Lifshitz transition ($\kappa \to \kappa_c$), pertinent to C-IC transition in a grand-canonical ensemble, is replaced by a linear one.
A similar situation has been described earlier
\cite{vekua} in the study of magnetization process of isolated 1D Fermi systems
at a fixed particle number, and, more recently, in the context of 
standard ladders away from 1/2-filling \cite{caza}.

\subsection{$\vrho{\bf >1 ~(n>0)}$}
 \label{r-grt-1}

\begin{figure}[H]
\centering
\includegraphics[width=3.3in]{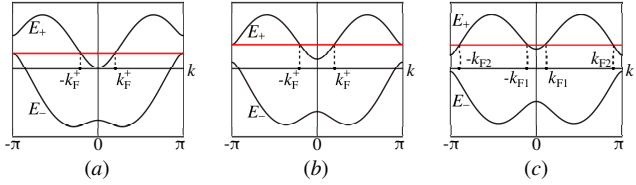}
\caption{\footnotesize  Case $\rho > 1$. (a) Lifshitz transition at $\kappa = \kappa_{c1}$ between
4FP-1 to 2FP-1 phases;
(b) Lifshitz transition at $\kappa = \kappa_{c2}$ between 2FP-1 to 4FP-2 phases; (c)  4FP-2 metallic phase at
$\kappa < \kappa_{c2}$.}
\label{224}
\end{figure}

Here again we start from the four-Fermi-point metallic state  which we now label as the
4FP-1 phase. At $n>0$ we have $p_+> p_-$.
On decreasing $\kappa$ 
the first Lifshitz point $\kappa = \kappa_{c1}$ is reached, Fig.\ref{224}a,
at which $p_- = 0$ and the band $E^- (k)$ gets completely filled. The
partial filling of the upper band $E^+ (k)$ within the interval
$|k| < k^+ _F$ reflects the existence 
of a finite particle doping ($n = \rho -1 > 0$). 
At the transition the number of Fermi points changes from four to two.
We will label this phase as 2FP-1.
We note that under the substitutions 
$p_+ \leftrightarrow p_-$, $M  \leftrightarrow|m|$, $n \leftrightarrow - n$, $\mu  \leftrightarrow- \mu$
the already discussed 4FP-2FP Lifshitz transition at $\rho < 1$ exactly maps to the  4FP1-2FP1  transition  
at $\rho > 1$. 
Therefore the critical values of $\kappa$ and $\mu$ are given by
$
\kappa_{c1} = \left( Q_M + |m|\right)/2$,
$\mu_{c1} =  \left(  Q_M - |m| \right)/2$,
with
$
Q_M = \sqrt{(\pi n)^2 W^2  + M^2},
$
and,  in the vicinity of the Lifshitz point, the flux dependence of the momentum 
 $p_- (\kappa)$  is given by
\bea
&&p_- =\frac{|m|\pi n} {A(|m|+Q_M)}
\left[ \sqrt{1 + (\zeta/\zeta_1)} - 1 \right], \label{P-int1}\\
&&\zeta_1 = \frac{1}{2A} \frac{|m|(\pi n W)^2}{Q_M (|m|+Q_M)^2} \ll 1
\nn
\eea
where $\zeta_1 = (\kappa - \kappa_{c1} )/\kappa_{c1}$~$(0<\zeta \ll 1)$.
As in the $\rho<1$ case, in (\ref{P-int1})
there are no restrictions on $n$ except for
$n \ll 1$.
The smallness of $\zeta_1$ implies the existence of the linear 
($p_- \sim \zeta$) and square-root ($p_- \sim \sqrt{\zeta}$) asymptotics in the regions
 $\zeta \ll \zeta_1$ and  $\zeta_1 \ll \zeta \ll 1$, respectively.
\medskip

With $k^+ _F$ fixed at the value $k^+ _F= \pi n > 0$, on decreasing $\kappa$ with $m$ kept finite
one finds the second
Lifshitz point $\kappa = \kappa_{c2} < \kappa_{c1}$ at which
the chemical potential 
reaches the value $\mu = E^+ (\pi)$ (see Fig.\ref{224}b).  The spectrum  displays a re-entrant transition to
another 4FP metallic phase which we denote by 4FP-2 (Fig.\ref{224}c).
One finds that
$
\kappa_{c2} = \left( Q_M - |m|\right)/2$,
$\mu_{c2} =\left(  Q_M + |m| \right)/2$.
This critical  point belongs to
the region $(M+|m|)/2 < \kappa \ll W $
if the density of doped particles 
$n > n^*$, where
$
n^* = ({2}/{\pi W}) \sqrt{|m|(M + |m|)}.
$
At a lower density, $n < n^*$, the  point $\kappa_{c2}$
is located at smaller values of $\kappa$: 
$
(M-|m|)/{2} < \kappa < (M+|m|)/{2},
$
with $p_2 \to 0$ as $\kappa \to \kappa_{c2} - 0$.
\vskip 0.2cm

As shown in  Fig.\ref{224}c, in the 4FP-2 phase
the four Fermi points
$\pm k_{F1}$ and $\pm k_{F2}$ are all located in the upper band $E^+ (k)$
and are subject to the condition $k^+ _{F1} + (\pi - k^+ _{F2}) = \pi n$.
They can be parametrized as
 $k_{F1} = \pi n - p_2$, $k_{F2} = \pi - p_2$ ($0 < p_2 \ll 1)$.
At the transition ($\kappa = \kappa_{c2}$)  $p_2 = 0$. 
Following the same procedure as before 
we obtain
\bea
&& p_2 =\frac{(\pi n) |m|}{A(Q_M - |m|)}\left[
\sqrt{1 +\left(\zeta/\zeta_2 \right) }  -1 
\right], \label{P-int2}\\
&&\zeta_2 = \frac{|m|(\pi n W)^2 }{2A Q_M (Q_M - |m|)^2}
\nn
\eea
where $\zeta_2 = (\kappa_{c2} - \kappa)/\kappa_{c2} > 0 (~\zeta_2 \ll 1)$.

\begin{figure}[H]
\centering
\includegraphics[width=3.2in]{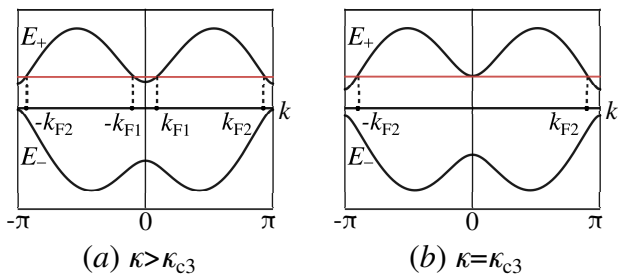}
\caption{\footnotesize Lifshitz transition from 4FP-2 to 2FP-2 at $\rho>1$.}
\label{4Fp-2Fp}
\end{figure}

 Further decreasing $\kappa$  in the 4FP-2 phase one passes to
the region
$
\kappa < (M-|m|)/2 .  
$
The spectrum is shown in Fig.\ref{4Fp-2Fp}a.
The Fermi momenta $k_{F1}$ and  $k_{F2}$   are parametrized as
$k_{F1} = p_1$, $\pi - k_{F2} = \pi n - p_1$.
$p_1$ and $\mu$ are
determined by the equations
\bea
\sqrt{p^2 _1 v^2 _F + M^2} &=& \mu + \kappa, \nn\\
\sqrt{(\pi n - p_1)^2 v^2 _F + m^2} &=& \mu - \kappa
\label{eqs12}
\eea
There exists the third
Lifshitz transition at $\kappa = \kappa_{c3}$ 
at which $\mu = E^+ (0)$, $p_1 = 0$; see Fig.\ref{4Fp-2Fp}b. 
We find that
$
\kappa_{c3} =  \left( M - Q_m \right)/2$, ~
$\mu_{c3}  = \left( M + Q_m \right)/2$.
The requirement $\kappa_{c3} > 0$ is ensured by the condition (\ref{n<}) which sets
the upper limit for $n$.
From Eqs.(\ref{eqs12}) one obtains the quadratic equation
valid at $\kappa > \kappa_{c3}$:
\bea
\left[  1 - \frac{(\pi n)^2 W^2}{4\kappa^2}  \right] (p _1 v_F)^2
- (\pi n W)  \left(  1 + \frac{\mu_{c3} \kappa_{c3}}{\kappa^2} \right) (p_1 v_F)\nn\\
+ \left(\frac{\mu^2 _{c3}}{\kappa^2} - 1\right) \left( \kappa^2 - \kappa^2 _{c3}  \right) = 0
~~~~~\label{exact-equation2}
\eea
Contrary to all previous cases, a physically acceptable solution of Eq.(\ref{exact-equation2})
only exists if $\pi n W > 2\kappa$. 
This
brings the allowed
doped particle density $n$ to 
the interval
\be
\frac{M^2 - m^2}{2MW} < \pi n < \frac{\sqrt{M^2 - m^2}}{W} \label{n-range}
\ee
Taking the limit $M \gg |m|$, one then concludes that the transition at $\kappa = \kappa_{c3}$
and the onset of the 2FP-2 phase is only possible at   $n \sim M /W$.
Choosing $|1 - (\pi n W)^2/\kappa_{c3}^2| \sim 1$, $Q_m \gtrsim M$, $\pi n W \sim M$
we find that
\be
p_1 (\zeta_3) \sim \sqrt{\zeta_0 + \zeta_3} - \sqrt{\zeta_0}, ~
\zeta_0 \sim \frac{(\pi n)^2 W^2 M}{(M-Q_m)^2 Q_m} 
\ee
where $\zeta_3 = (\kappa - \kappa_{c3})/\kappa_{c3}$.
Since $\zeta_0 \sim 1$, only the linear dependence of the momentum $p_1$ on $\zeta_3$
at $\zeta_3 \ll 1$ is realized. There is no
crossover to a square-root asymptotics   in this case. 

Thus, contrary to the case $\rho<1$ in which only one Lifshitz transition takes place,
at $\rho > 1$ and $m \neq 0$, on increasing the flux towards the value 1/2 (decreasing $\kappa$ towards zero)
the spectrum of the system
undergoes 
three consecutive Lifshitz transitions 
\bea
\kappa = \kappa_{c1}:  && \textrm{4FP-1 metal} ~\to~ \textrm{2FP-1 metal} \nn\\
\kappa = \kappa_{c2}:  && \textrm{2FP-1 metal} ~\to~ \textrm{4FP-2 metal} \nn\\
\kappa = \kappa_{c3}:  && \textrm{4FP-2 metal} ~\to~ \textrm{2FP-2 metal}
\nn
\eea
where $\kappa_{c1} > \kappa_{c2} > \kappa_{c3}$. In the vicinity of each transition, the behavior of the orbital current  at $\rho>1$
is similar to the already considered case $\rho < 1$.
\begin{figure}[H]
\centering
\includegraphics[width=1.7in]{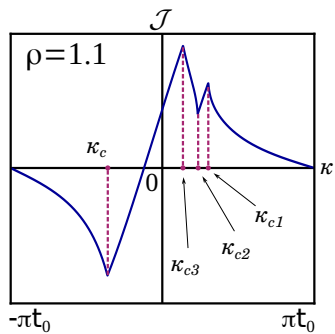}
\caption{\footnotesize Cusps in the flux dependence of the orbital current at 
$\rho > 1$.}
\label{cusps}
\end{figure}
Fig.\ref{cusps} shows the flux dependence of the orbital current of a particle-doped ladder ($\rho > 1$)
for both signs of $\kappa = 2\pi t_0 (1/2 - f)$. Note that due to the symmetry property (\ref{-symA}), 
in Fig.\ref{cusps} the region $\kappa < 0$ with $\rho = 1.1$ can be mapped to the region $\kappa > 0$
with $\rho = 0.9$. Therefore Fig.\ref{cusps} actually demonstrates the clear asymmetry  in the
flux dependence of the current for particle and hole doped samples (three Lifshitz transitions versus one).

\begin{figure}[H]
\centering
\includegraphics[width=1.7in]{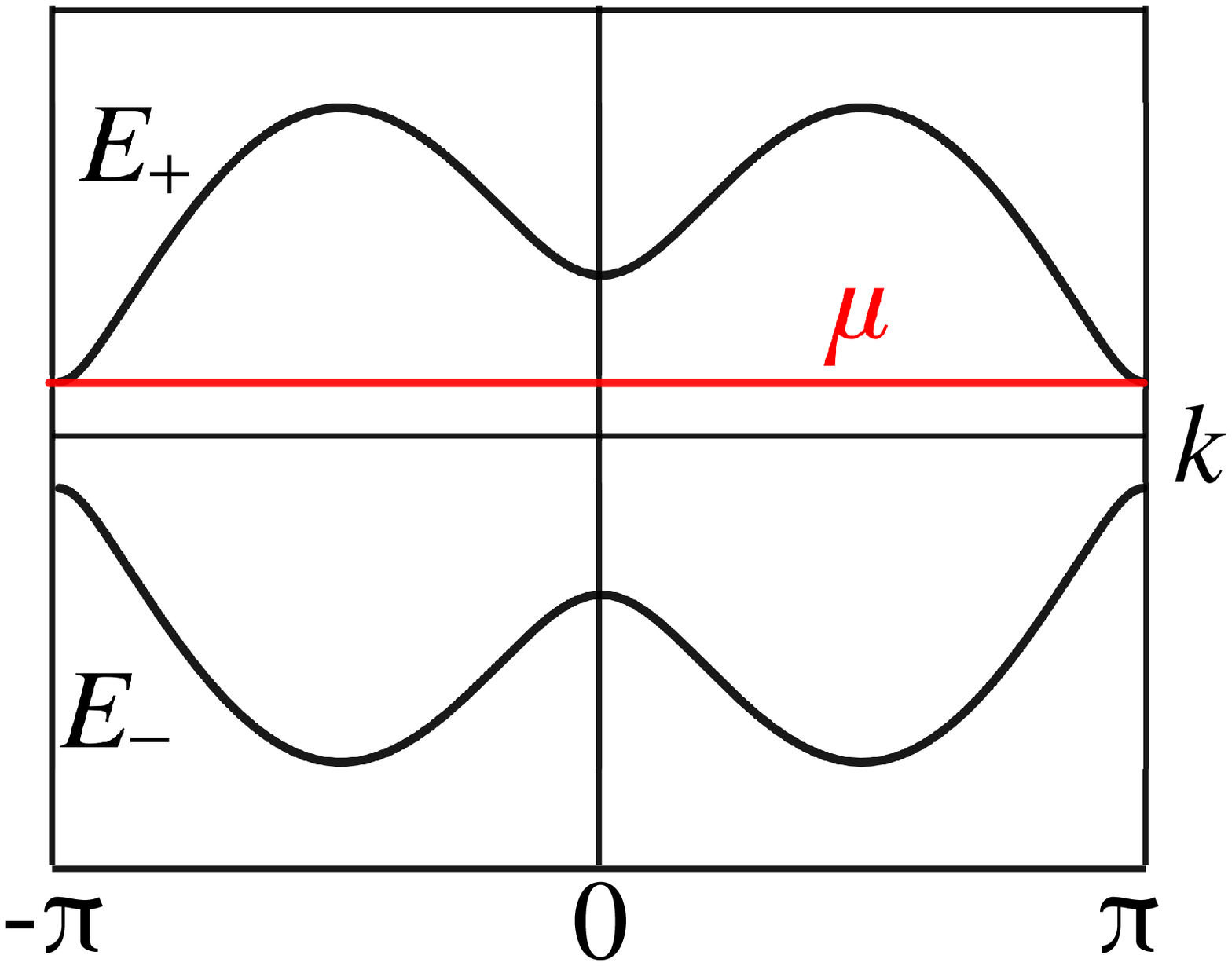}
\caption{\footnotesize $\rho \to 1+$, $m\neq 0$} 
\label{ins-2}
\end{figure}

In the 2FP-2 metallic phase the Fermi momentum $k^+ _{F2}$ is fixed by the condition
$k^+ _{F2} = \pi (1-n)$. 
The chemical potential is given by
$
\mu = \kappa + \sqrt{(\pi n)^2 + m^2}.
$
At $n \to 0^+$, $ \rho \to 1^+$, the band $E^+ (k)$ becomes empty
with the chemical potential attached to its bottom:
\be
\lim_{n\to 0} \mu(\kappa) = \kappa + |m|
\label{mu-lim-1+}
\ee
Comparing Figs.\ref{1+A}a and \ref{ins-2}, and formulas (\ref{mu-n-0}) and (\ref{mu-lim-1+}), respectivel;y,
we observe that at $\rho \to 1$ the chemical potential 
displays a discontinuity :
\be
\mu_{\pm} (\kappa) \equiv \lim_{\rho \to 1^{\pm}} \mu (\kappa)= \kappa \pm |m| \label{mu-jump}
\ee
Since the classic paper by Lieb and Wu \cite{LW} this fact is regarded 
as a direct manifestation of the insulating nature of the ground state of a system.
In fact, for a system with given particle number $N$ with the ground state energy
$E_0 (N)$, the chemical potentials $\mu_{\pm}$ are defined
as $\mu_+ = E_0 (N+1) - E_0 (N) $ and $\mu_- = E_0 (N) - E_0 (N-1) $.
The finite difference $\mu_+ - \mu_- = 2|m| > 0$ then defines the spectral gap of the
insulating state.
\medskip

We complete the discussion by noting that the limit $\rho \to 1$ is unambiguous
when the $k\sim \pi$ fermions are massless: $m = 0$. In this case the ground state of the system is always
metallic, characterised by a single Dirac point in the spectrum.

\section{Conclusion and outlook}\label{concl}

In this paper we have shown that a 
noninteracting model of spinless fermions on an asymmetric triangular ladder 
exhibits interesting properties originating from  the interplay of geometrical frustration of the lattice and
the external magnetic flux. The main results of our study is applicable to a broader class of frustrated ladders
which can be unified into asymmetric Creutz model depicted in Fig.\ref{creutz}, the TFL being the 
most frustrated member of this family.
Contrary to the standard (rectangular) ladder, in a TFL  the $k \to \pi - k$
particle-hole symmetry of the  two-band spectrum is absent. At a weak interchain hopping
qualitative differences between the two systems become evident when the flux per the diatomic plaquette
is close to  flux quantum. In this regime, the removal of  the double degeneracy of the spectral gap 
leads to the appearance of two groups of low-energy Dirac-like excitations with different masses.
Even for a nearly half-filled spectrum this leads not to a single metal-insulator  transition as in the standard ladder, but to a cascade of flux-induced 
Lifshitz transitions that separate phases with the number of Fermi points equal to 4, 2, 1 or 0, the latter case corresponding to
the band insulator phase.
The transitions take place when the flux is varied.
At the critical points the orbital current 
and its derivative, the charge stiffness,
display singular behavior: universal square-root singularities of the "commensurate-incommensurate" transition type when the chemical potential $\mu$ 
is constant, or cusps when a system is considered at a fixed particle density $\rho$. We have shown that the TFL  displays a
generic property that follows from the absence of $k\to \pi - k$ particle-hole symmetry-- different patterns of Lifshitz transitions  at a particle ($\rho > 1$) and hole ($\rho<1$) doping.

Remarkably, for a translationally
invariant TFL ($m=t_1 - t_2 = 0$) and the flux close to the value $f=1/2$, the excitations  close to the Brillouin zone boundary ($k=\pi$)
become gapless and, in the low-energy limit, transform to a massless Dirac fermion. Such system is very susceptible  to inter-particle correlations. 
At sufficiently weak interaction the semi-metallic phase of the TFL will transform to a Tomonaga-Luttinger liquid.
However, on increasing interaction one expects
the onset of strongly correlated phases formed due to the 
two-particle  interchain processes that become relevant in the renormalization-group sense.
As long as the mass gap $|m_s|$  dynamically generated in the strong-coupling phase of the interacting massless
model is much smaller than $|m|$, the single-particle branch of excitations still remains well defined up small renormalizations of
the mass $m$ and velocity $v_F$ caused by interactions. With such renormalizations taken into account, the results obtained in the present work 
are applicable to the interacting case as well.
As in the non-interacting case, the ground state of the model would be  characterized by
an explicitly broken parity ($x\to - x$) with patterns of dimerization of the zigzag bonds  depending on the sign of $m$.
The most interesting regime is expected to take place in the range of parameters when the two masses, $m_s$ and $m$, become of the same
order.
These and related issues will be addressed in a separate forthcoming publication \cite{cooper}.

From the theoretical point of view,
the solution of the two-leg TFL  is expected to represent 
a building block for solving frustrated flux-ladders with a larger
number of chains. It is very interesting to formulate a field-theoretical approach to study non-perturbative effects in a weakly coupled multi-chain ladders
using methods of Abelian and non-Abelian bosonization \cite{GNT}.
A very different and  complicated is the problem of frustrated ladders with a staggered flux. Here one faces a completely different physics
already at the level of non-interacting fermions.
One of the most challenging issues here is a chiral asymmetry of the spectrum and breakdown of Lorentz invariance.

Recent advances in experimental realization of a Creutz flux ladder for ultracold fermionic atoms in a resonantly driven 1D optical lattice \cite{kang}
make us believe that the results obtained in the present paper admit their experimental verification. 

After completion of this paper we become aware of the recent work \cite{giam-2022} in which a bosonic two-chain triangulat ladder in the 
presence of an artificial flux is considered. Even though the physics of bosonic and fermionic
flux ladders is quite different,  the two systems share the important common feature -- the enhanced role of geometrical frustration
of the underlying lattice at the values of the flux close to a flux quantum per elementary plaquette.

\medskip

{\bf Acknowledgements}
\medskip

We thank  Marcello Dalmonte for reading the manuscript and making a number of important suggestions.
We thank him,
Titas Chanda, Pierre Fromholz, Emanuele Tirrito and Mikheil Tsitsishvili
for fruitful discussions and cooperation in related projects. We are also grateful to 
George Japaridze and Oleg Starykh for interesting conversations on the effects of frustration in one-dimensional quantum systems.
The authors  acknowledge the support from the Shota Rustaveli National Science
Foundation of Georgia, SRNSF, Grant No. FR-19-11872.

\end{document}